\documentclass[12pt,aps,eqsecnum,amsfonts,showkeys]{revtex4}

\usepackage{theorem}
\theorembodyfont{\upshape}

\input epsf

\usepackage{graphicx}

\usepackage{bm}

\newcommand{\beq}{\begin{equation}}
\newcommand{\eeq}{\end{equation}}
\newcommand{\beqs}{\begin{eqnarray}}
\newcommand{\eeqs}{\end{eqnarray}}

\begin{document}

\pagestyle{myheadings}

\title{Essential electronic properties on stage-1 Li/Li+-graphite-intercalation compounds for different concentrations}

\author{Wei-Bang Li}
\email{weibang1108@gmail.com}

\author{Shih-Yang Lin}
\email{sylin1985@gmail.com}

\author{Ngoc Thanh Thuy Tran}
\email{thuytran74vn@gmail.com}

\author{Kuang-I Lin}
\email{kilin@mail.ncku.edu.tw}

\author{Ming-Fa Lin}
\email{mflin@mail.ncku.edu.tw}

\affiliation{ \ Department of Physics \\
National Cheng Kung University \\
Tainan 70101, Taiwan}

\begin{abstract}

We use first-principles calculation within the density functional theory (DFT) to explore the electronic properties on stage-1 Li- and Li+-graphite-intercalation compounds (GIC) for different concentrations, LiCx/Li+Cx with x= 6,12,18,24,32 and 36. The essential properties, e.g. geometric structures, band structures and spatial charge distributions are determined by the hybridization of orbitals, the main focus of our works. The band structures/density of states/spatial charge distribution display that the Li-GIC possesses blue shift of fermi energy and just like metals, but the Li+-GIC still preserves as original graphite or so-call semimetal possessing the same densities of free electrons and holes. According to these properties, we find that there exists weak but significant van der Waals interactions between interlayer of graphite, and 2s-2pz hybridization between Li and C. There scarcely exists strong interactions between Li+-C. The dominant interaction between the Li and C is 2s-2pz orbital-orbital couple; the orbital-orbital couple is not significant in Li+ and C case but the dipole-diploe couple.

\end{abstract}

\maketitle

\section{Introduction}

Green energy is one of the most important issues in the whole world. The limited resource, such as fossil fuel, will be consume eventually. Besides that, the concomitant environmental pollution is also inevitable. Therefore, it gives rise to the efficient storage of electrical energy. Apart from the resources on earth, the solar power is the most potential energy for human. Years passed, to dealing with the problems of energetic storage, the rechargeable batteries emerged, which possess anodes, cathodes and electrolytes. Many previous theoretical and experimental studies are focused on lithium batteries and lithium-ion batteries because of the low costs, high safety and long cycle life. For the most part, the anodes are composed of carbon and non-carbon materials. The former is the well-known graphite, and the latter can be lithium, sodium or other ionic clusters. This kind of anodes are almost performed as graphite-intercalation compounds (GIC), that is to say, the lithium atoms (or other atoms) are intercalated into the graphite layers.
Graphite with layered structures is easily intercalated by alkali-metal atoms. The carbon-layered system is purely composed of the hexagonal symmetric lattice, in which there exists week but significant van der Waals (vdW) interactions. The vdW interactions modify the low-lying energy band structures and dominate the essential physical properties of graphite of GIC. The electronic properties strongly depend on the stacking configurations the graphitic layers perform. In general, there are three well known kinds of stacking configurations: AA (simple hexagonal), AB (Bernal), ABC (rhombohedral) \cite{001,002,003}. The previous studies\cite{004} reveal that the Li-GIC and A-GIC (A stands for alkali metal, A= Na, K, Rb, Cs) present the different structures in stage-1, respectively, LiC6 and AC8. Also, the AA-stacking configuration is the most stable one for Li-GIC on stage-1 type\cite{005}, whereas other stage-n types (n=2,3,4…) form the AB-stacking configuration. The comparisons between Li-GIC and A-GIC are widely discussed in the past years, but few theoretical and experimental researches on Li-ion-GIC and alkali-ion-GIC are explored.
In this work, we mainly focus on the orbital hybridizations in lithium and lithium-ion GICs by presenting the essential structural and electronic properties. We consider the different concentrations on uniform situation of intercalations, i.e., this paper only covers the stage-1 type for different concentrations. In addition, the optimal adatom positon is located in the hollow site\citep{006,007,008,009} of hexagonal carbon layers. The first principle calculations\cite{010,011,012,013} are utilized to investigate the total ground state energies, optimal geometric structures, energy band structures, density of states and the spatial charge distributions. Moreover, the intercalation-induced conduction-electron densities can be predicted. The rich and unique phenomena in Li-GIC and Li+-GIC are expected to have the significant differences under the symmetric comparison with each other.

\section{Theoretical calculations}

The density functional theory (DFT)\cite{014,015,016,017} has been widely utilized for the many-electron systems in physics and chemistry for years. Specially, the Vienna ab initio simulation package (VASP)\cite{018} evaluates an approximate solution within the density functional theory by solving the Kohn-Sham equations. We use first-principle calculations with VASP in this paper. The VASP applies the Bloch’s theorem because it is suitable to deal with the problems of bulks with periodic boundary condition. The exchange-correlation energy depending on electron-electron interactions is calculated from Perdew-Burke-Ernzerhof functional under the generalized gradient approximation (GGA)\cite{019}. First, all the atoms in the relaxation process are adjusted to form the optimized structures with the lowest total ground energy. The spatial charge density could be solved by numerical self-consistent scheme. Furthermore, we can use the above results to investigate the fundamental physical properties. The first Brillouin zone is sampled in a Gamma scheme by $10\times 10\times 10$ k-points meshes for optimization and $30\times 30\times 30$ k-points meshes for electronic structures. The energy convergence is set to be $10^{-5}$eV for two simulation steps. The maximum Hellmann-Feynman force on each atom is less than 0.01 eV/$\dot{A}$. The details of calculation processes are in the flowchart [Fig. 1].

\section{Results and discussions}
\subsection{Geometric structures}

The geometric symmetries of graphite are diversified by the chemical intercalation. The Li/Li+ can be easily intercalated into the interlayer spacing because of the week but significant van der Waals interactions. There are three frequent types of absorptions positions, hollow site, top site and bridge site, for intercalant atoms. The hollow-site position possesses the lowest ground energies, that is to say, the hollow-site position is the most stable geometric configuration\cite{020,021} The pristine structures of LiCx/Li+Cx display the same stacking type, stage-1, but distinct concentrations, listed in Table I. The changes between interlayer distances of Li-GIC/Li+-GIC are very different from each other.
For atom cases, the interlayer distances of LiC6, LiC12, LiC18, LiC24, LiC32 and LiC36 are 3.815 $\dot{A}$, 3.863 $\dot{A}$, 3.924 $\dot{A}$, 3.954 $\dot{A}$, 3.987 $\dot{A}$ and 3.964 $\dot{A}$, respectively; for ion cases, the interlayer distances of Li+C6, Li+C12, Li+C18, Li+C24, Li+C32 and Li+C36 are 3.082 $\dot{A}$, 3.283 $\dot{A}$, 3.374 $\dot{A}$, 3.451 $\dot{A}$, 3.510 $\dot{A}$ and 3.555 $\dot{A}$, respectively. Seen in Table I.

The heights of LiCx and Li+Cx decrease with the higher concentrations. Apparently, compared to the pristine graphite lattice with experimental interlayer distance of 3.550 $\dot{A}$\citep{022}, the height of former will be closed to that of pristine graphite when the concentration rises; in the contrary, the height of latter will be closed to that of pristine graphite lattice when the concentration declines [Fig. 2].

\subsection{Band structures and Density of states}

The pristine AA-stacking graphite possesses an unusual electronic structure, including the same amounts of free electrons and holes within -0.5 eV$\sim$0.5 eV according to the band structure and density of state [Fig.3], and obvious band overlap is revealed in the kz-dependent energy dispersion along $\Gamma A$. That is to say, the AA-stacking graphite behaves like semimetals. The low-lying energy bands (π bands) are dominated by the C-2pz orbitals; the C-(2s, 2px, 2py) orbitals, generating the $\sigma$ bands, appears at E $\leq$ -3.0eV and strongly contribute to form the planar geometric structures. However, the electronic band structures exhibit very different changes after the intercalation of Li and Li+.

For intercalations of Li atom, the asymmetry of electron and hole bands becomes much notable. Apparently, the Fermi level presents the blue shift, compared to the pristine graphite, and does not intersects with any valence bands [Fig.4]. That is to say, the free conduction electrons are all created by the intercalations of Li atoms and replace the pristine carriers (both electron and hole). The blue shift of Fermi level is determined by the negative energy with the minimum density of state and the $E_{F}$ (E=0); the values are calculated to be, respectively, 1.800 eV, 1.326 eV, 1.138 eV, 0.991 eV, 0.882 eV and 0.810 eV for x=6, 12, 18, 24, 32 and 36, seen in Fig. 5. These results display that the C-2pz orbitals are easily affected by the Li-C bonds and sensitive to the concentrations of Li-atom intercalation.

For the intercalations of Li+ ions, it is quite different from the Li case. The band structures do not exhibit the apparent shift of Fermi level, estimated to be 0.088 eV, 0.042 eV, 0.036 eV, 0.026 eV 0.018 eV and 0.026 eV for x=6, 12, 18, 24, 32 and 36, seen in Fig. 5; that is, the Fermi level maintains the situation similar to the pristine graphite. Moreover, the initial $\sigma$ bands, displaying as shoulder structures in the orbit-projected density of states near -2.6 eV$\sim$-2.9 eV at $\Gamma$ point, present blue shift relatively to the Li ones; in the other words, the σ bands are deeper or more stable in the Li intercalations than in the Li+.

\subsection{Charge distributions and charge transfer}

The spatial charge distributions ($\rho$) and variations ($\Delta\rho$) of Li- and Li+-GICs are illustrated in Fig. 5 and Fig. 6. They are all shown on x-z plane and useful for understanding the chemical bonding changed after intercalations. For Li-atom intercalations, the $\rho$ and $\Delta\rho$ indicate that significant hybridizations of Li-C bonds with red and yellow colors in Fig. 6 (within the red dash frame) depend on the concentrations; the variations become obvious with the increase of concentrations, that is to say that the charge transfer is strongest in LiC6, but weakest in LiC36. Moreover, the variations are close to C atoms but not the Li atoms, that indicates the charge transfers from Li atom to C atom. Also, the rate of transferring mainly appear on the C atoms directly neighboring to Li atoms.

For Li+-ion intercalations, the variations present quite different characteristic. The $\Delta\rho$ between Li+ ions and C atoms are broader and lighter than the Li-atom cases with light green and yellow colors in Fig. 7 (within the red dash frame). In addition, the $\Delta\rho$ between the neighboring C atoms become obvious with the increase of concentrations with the light blue colors in Fig. 7 (within the black dash frame); also, the variations between Li+ ions and C atoms are slightly close to the former. That is to say, the charge transfers from C atoms to Li+ ions. It is similar to the Li-atom case that the carbons, not directly neighboring to the Li+ ions, are seldom affected. By the bader analysis in VASP calculations, the Li-GICs exhibit the charge transfers of for LiC6, LiC12, LiC18, LiC24, LiC32 and LiC36, respectively; the Li+-GICs exhibit the charge transfers of for Li+C6, Li+C12, Li+C18, Li+C24, Li+C32 and Li+C36, respectively. Seen in Table 2.

\section{Conclusions}

In summary, we present the current work from first-principles calculations within GGA method. The previous studies\cite{023,024} almost focus on Li-GICs but lack of Li+-GICs. Our works provide comparisons between these two cases and give the evidences of orbital hybridizations. We find that the changes of essential physical properties are very different between Li- and Li+-GICs, e.g., the interlayer distances, energy band structures, density of states and the spatial charge distributions, and the results clearly reveal that the Li-C bonds are generated from the 2s-2pz orbital hybridizations, it leads to the high charge transfers from Li to C about 0.86$\sim$0.88 as well. In the contrary, the variations are relatively week, the charge transfers, which Li+ obtains from C, merely range 0.163$\sim$0.191. The dominant effects between Li and C are obviously the orbital-orbital interactions; but between Li+ and C, the main effects might be the dipole-dipole interactions. The saturated electronic configurations of Li+ ions, similar to the inert gas helium [He], lead to low contributions of free electron carriers.

\bigskip

Acknowledgments:
This work is supported by the Hi-GEM Research Center and the Taiwan Ministry of Science and Technology under grant number MOST 108-2212-M-006-022-MY3, MOST 109-2811-M-006-505 and MOST 108-3017-F-006-003.

\newpage

\begin{figure}[tbp]
\par
\begin{center}
\leavevmode
\includegraphics[width=1.0\linewidth]{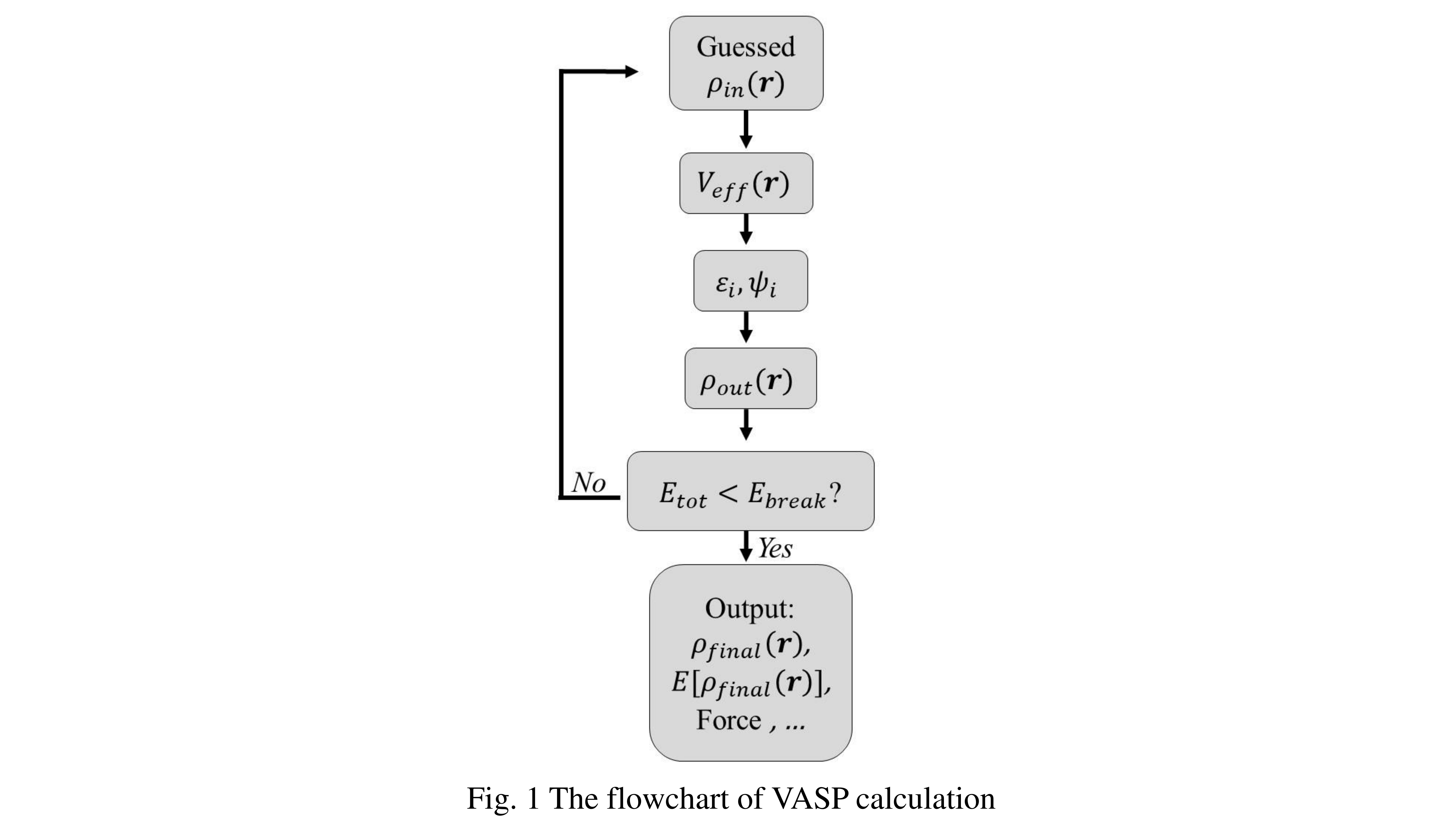}
\end{center}
\par
\end{figure}

\newpage

\begin{figure}[tbp]
\par
\begin{center}
\leavevmode
\includegraphics[width=1.0\linewidth]{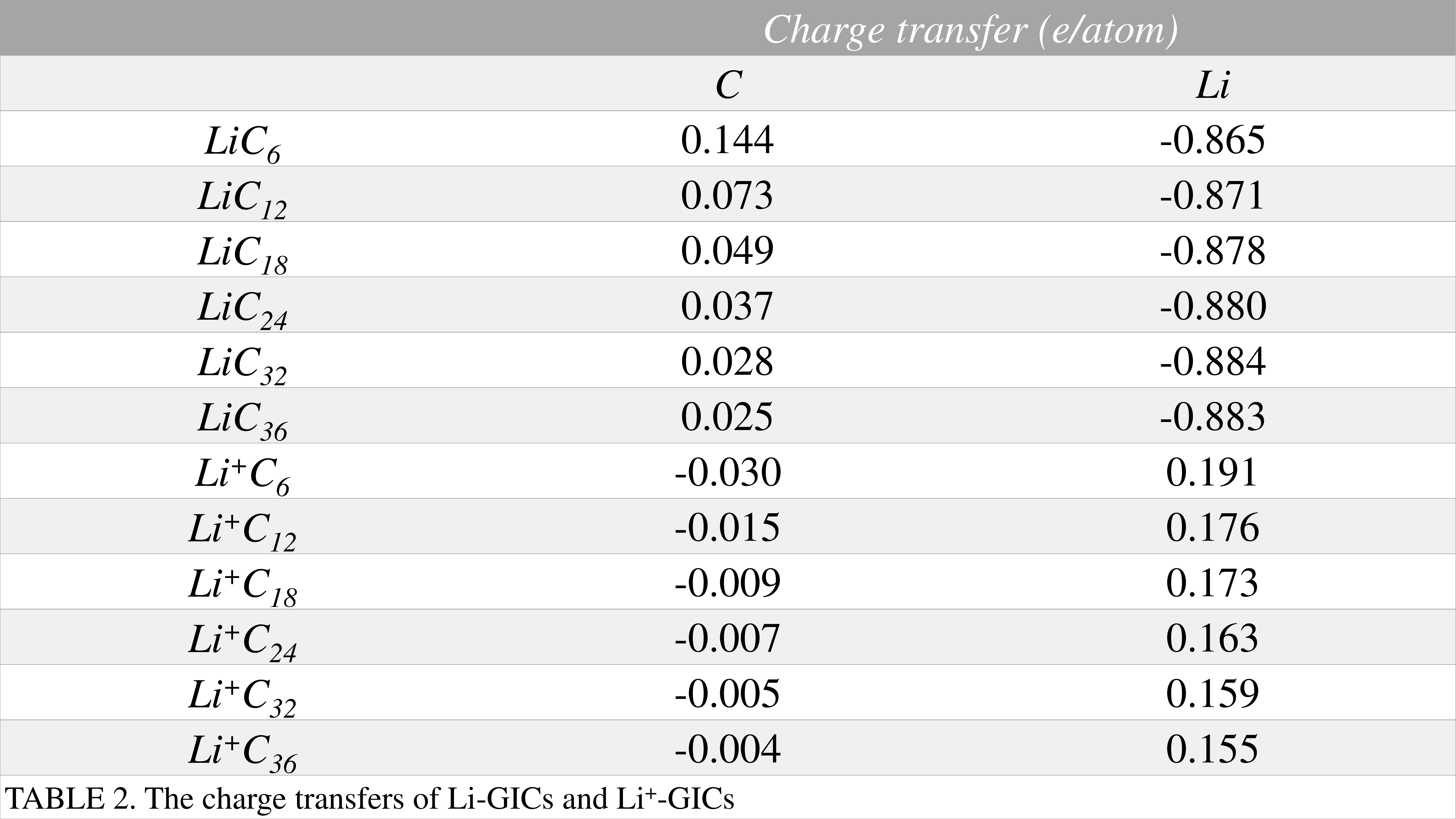}
\end{center}
\par
\end{figure}

\newpage

\begin{figure}[tbp]
\par
\begin{center}
\leavevmode
\includegraphics[width=1.0\linewidth]{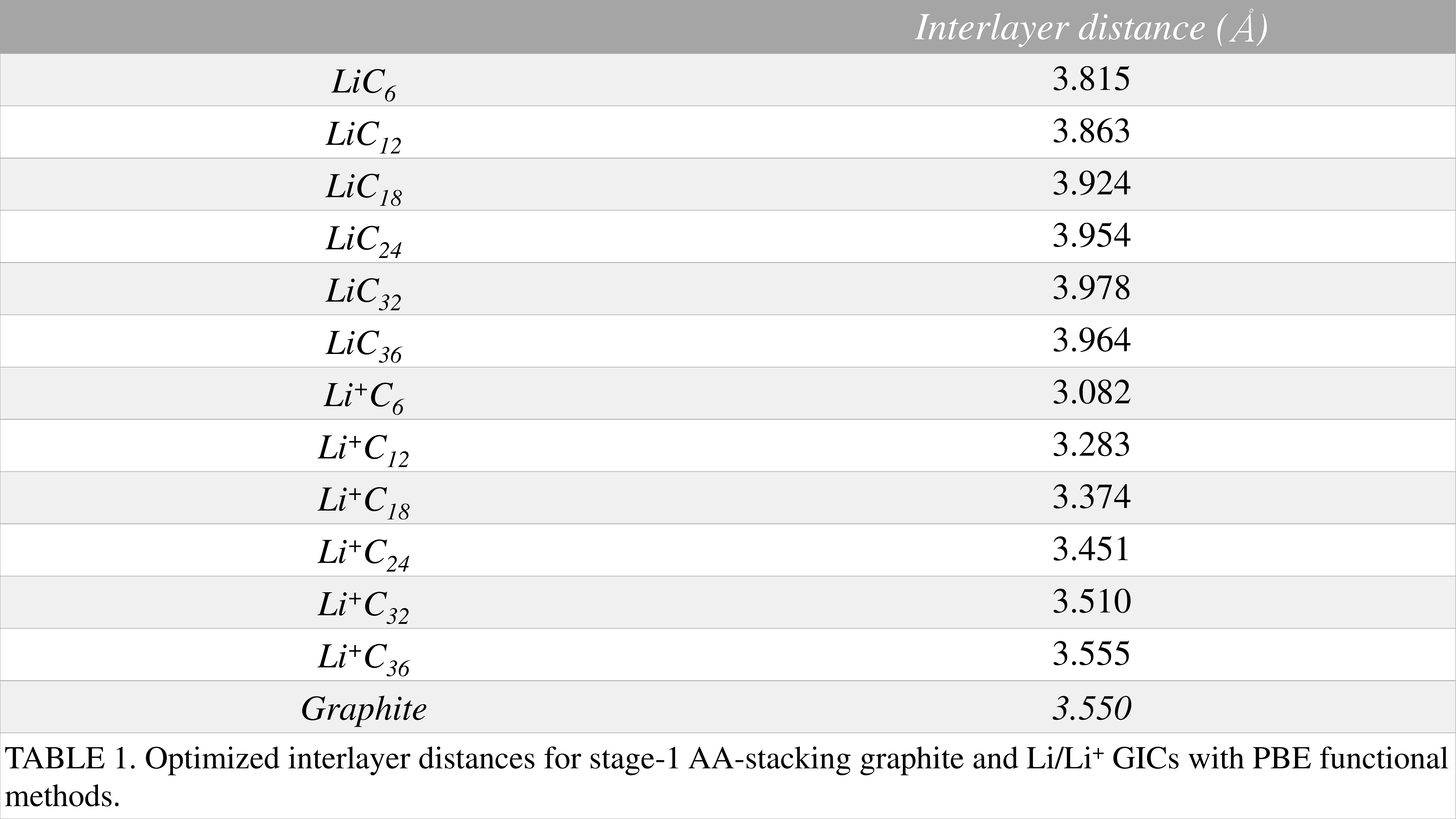}
\end{center}
\par
\end{figure}

\newpage

\begin{figure}[tbp]
\par
\begin{center}
\leavevmode
\includegraphics[width=1.0\linewidth]{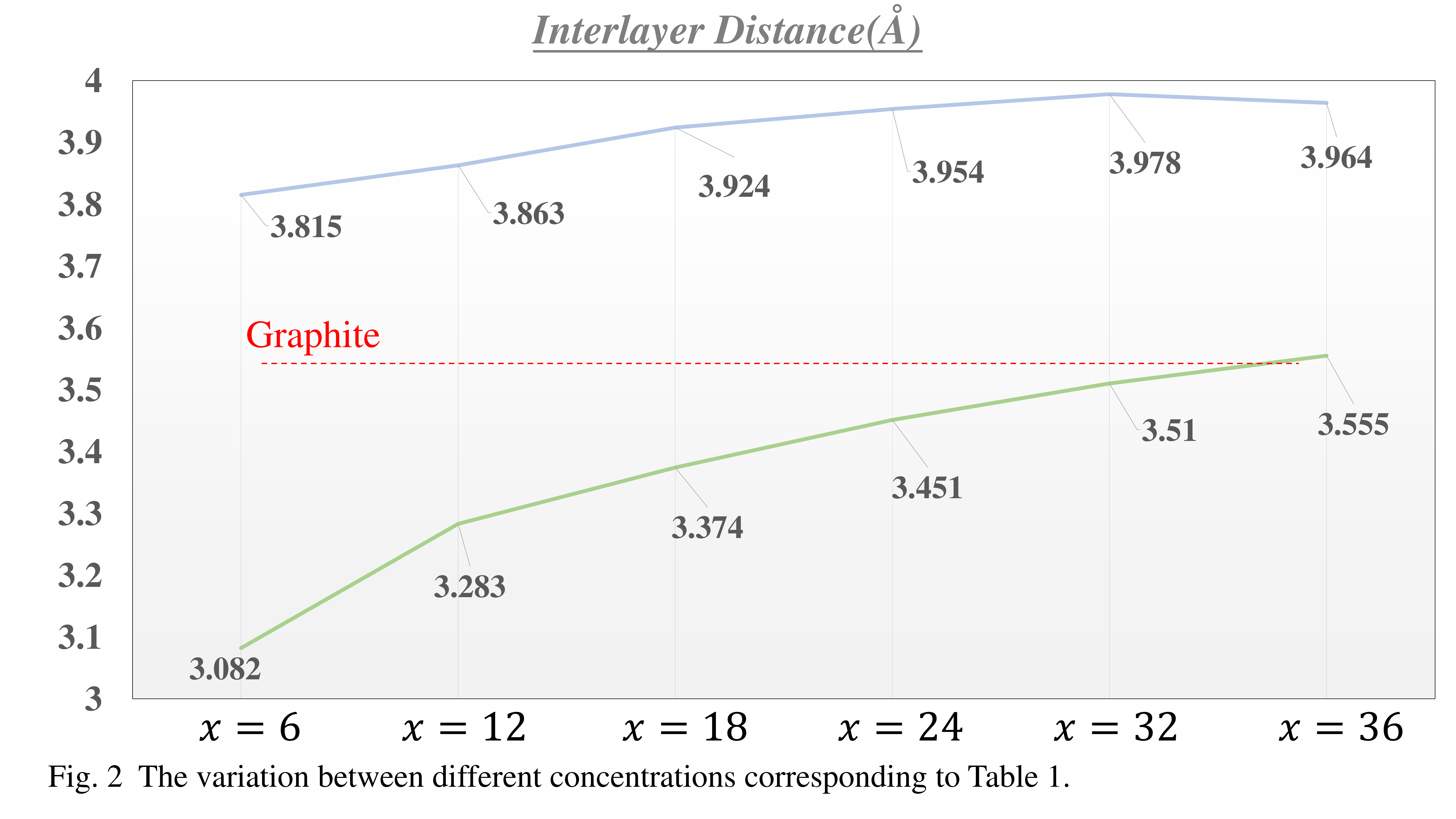}
\end{center}
\par
\end{figure}

\newpage

\begin{figure}[tbp]
\par
\begin{center}
\leavevmode
\includegraphics[width=1.0\linewidth]{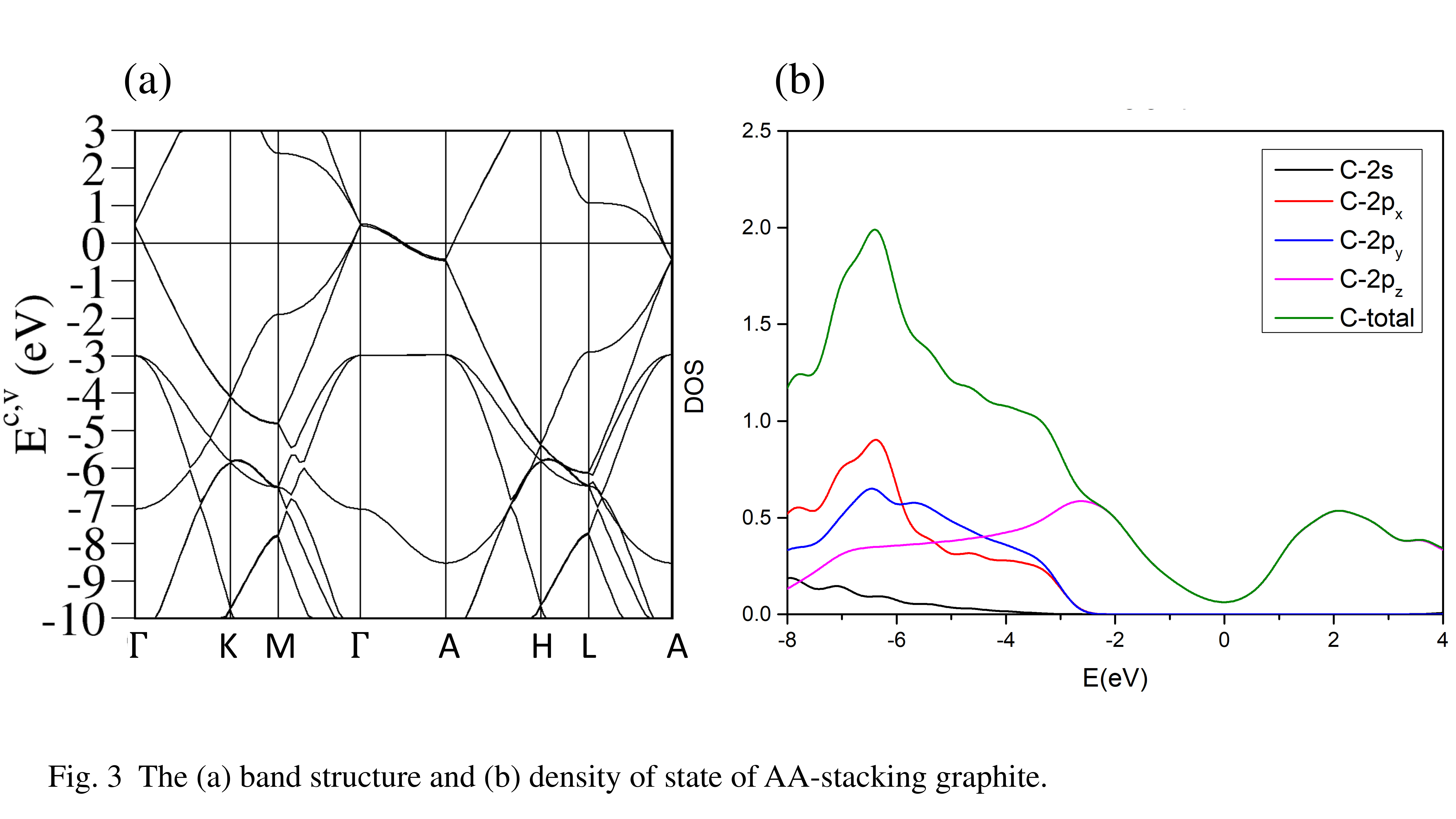}
\end{center}
\par
\end{figure}

\newpage

\begin{figure}[tbp]
\par
\begin{center}
\leavevmode
\includegraphics[width=1.0\linewidth]{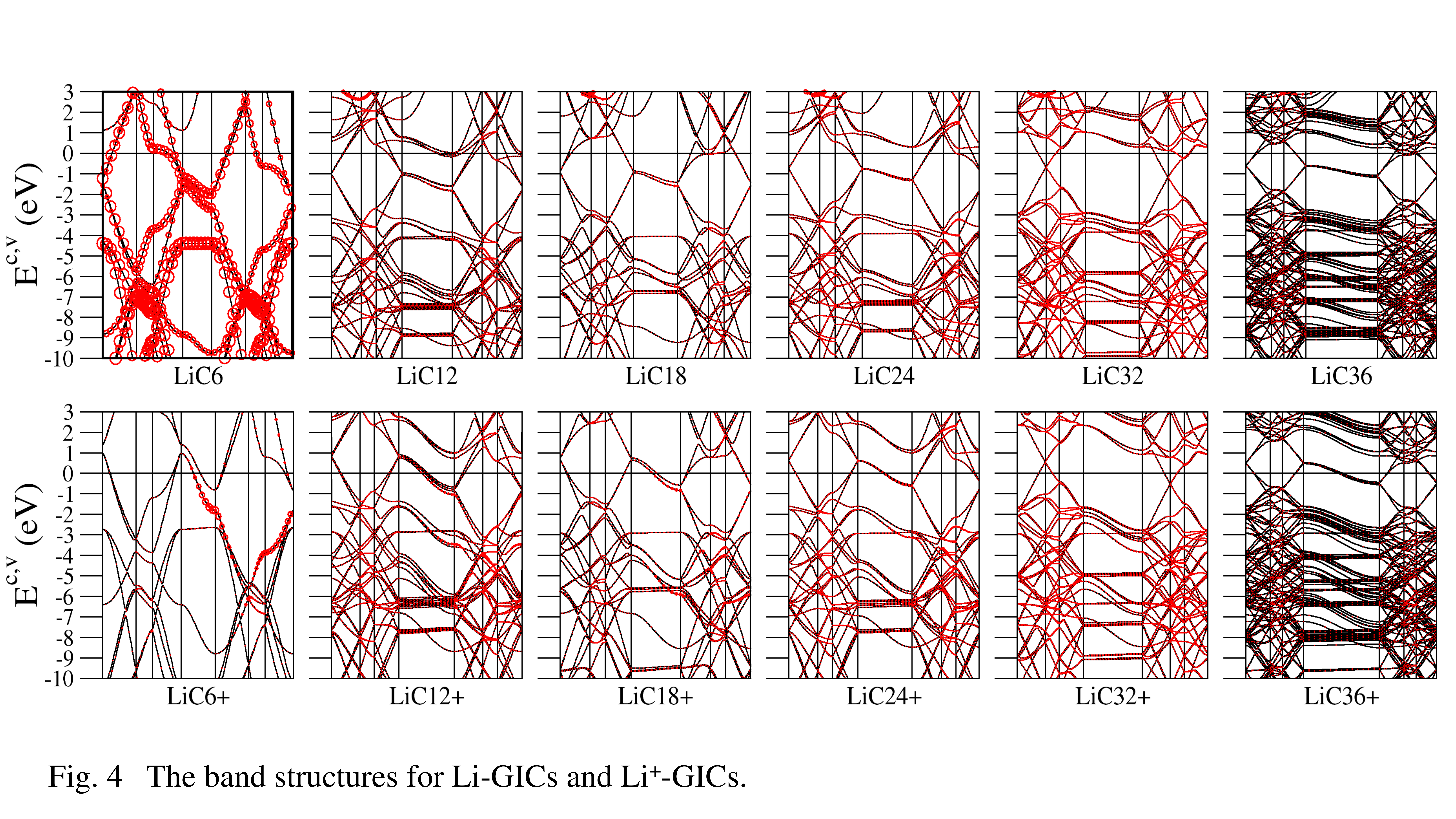}
\end{center}
\par
\end{figure}

\newpage

\begin{figure}[tbp]
\par
\begin{center}
\leavevmode
\includegraphics[width=1.0\linewidth]{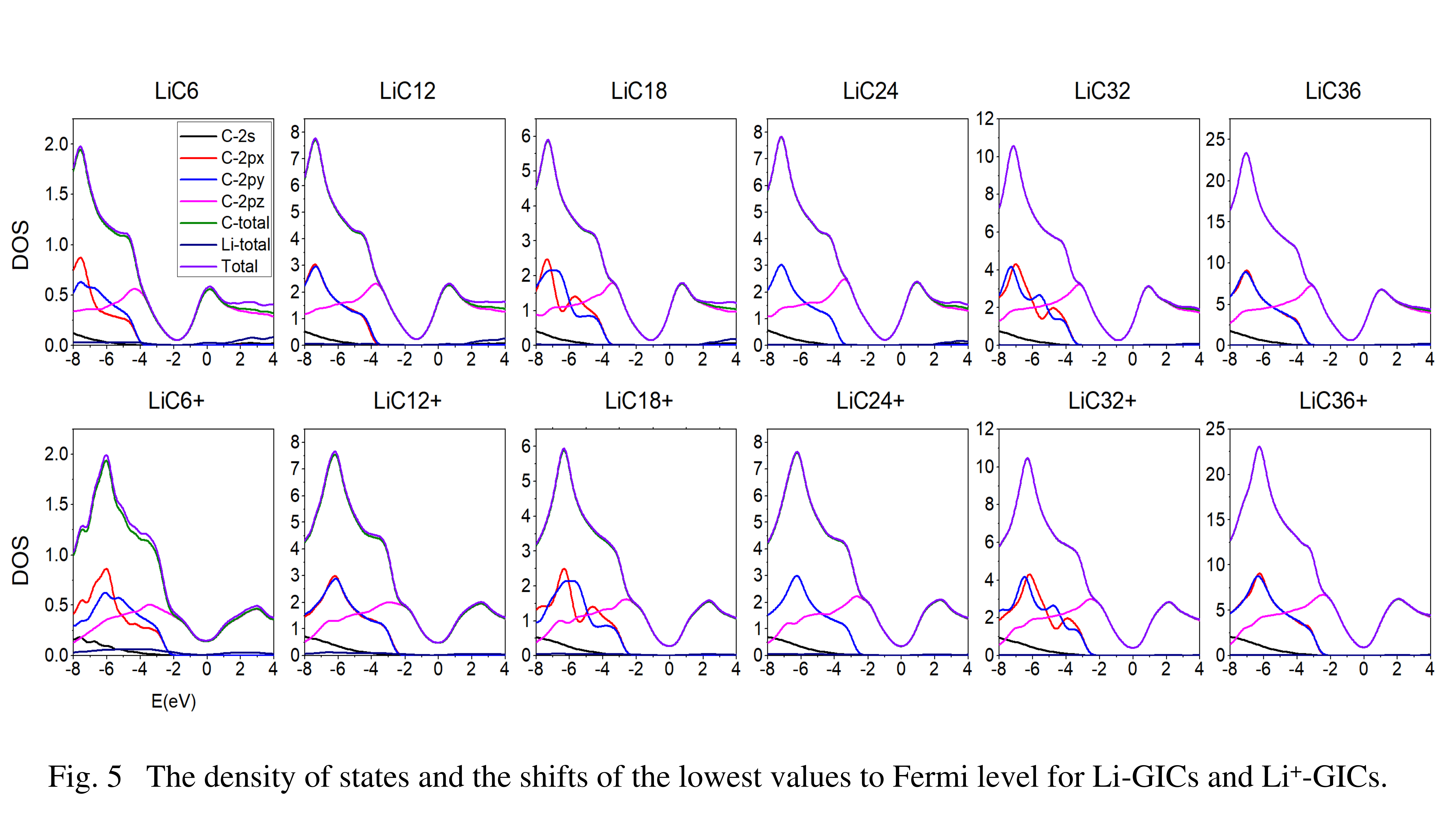}
\end{center}
\par
\end{figure}

\newpage

\begin{figure}[tbp]
\par
\begin{center}
\leavevmode
\includegraphics[width=1.0\linewidth]{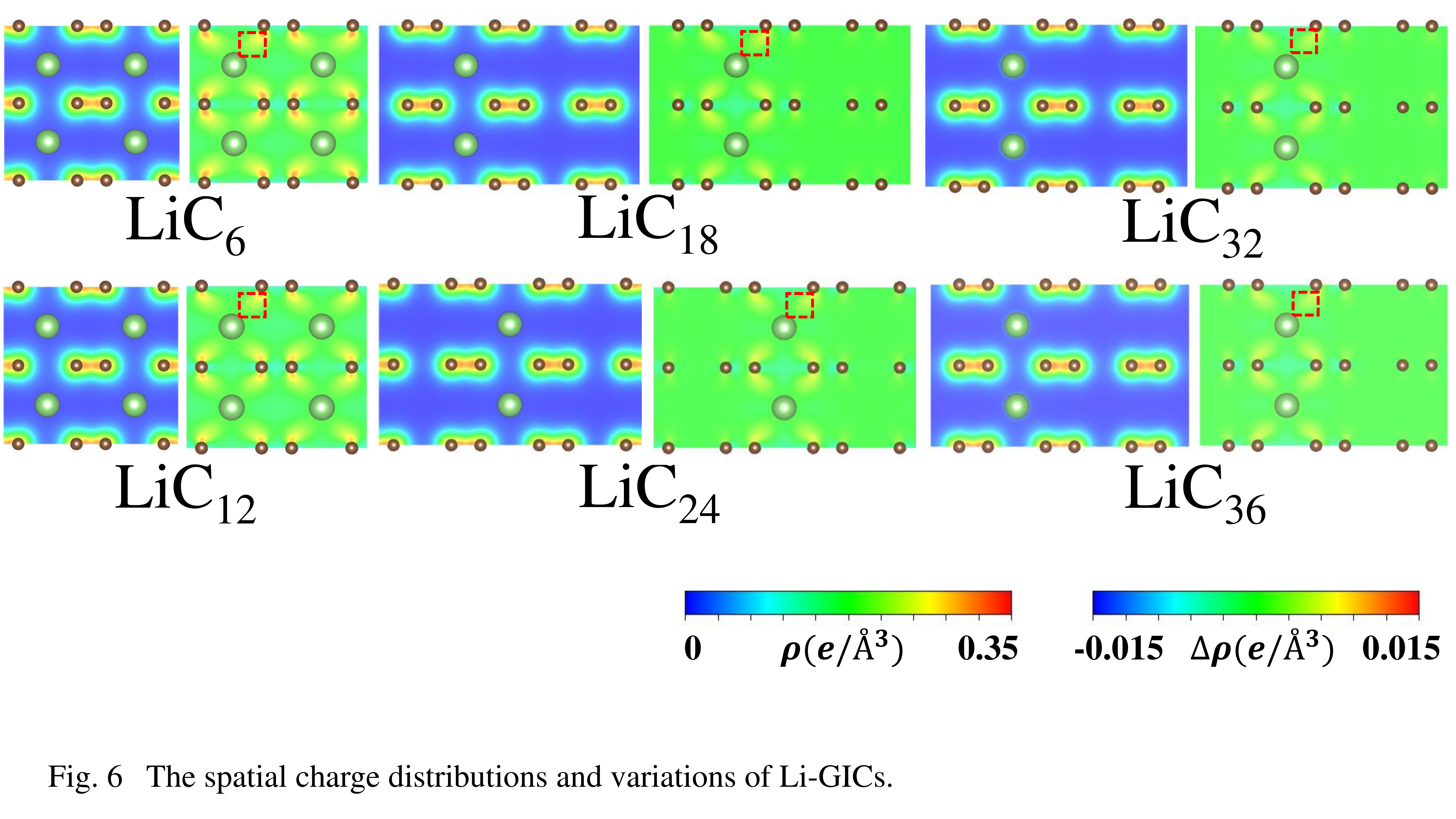}
\end{center}
\par
\end{figure}

\newpage

\begin{figure}[tbp]
\par
\begin{center}
\leavevmode
\includegraphics[width=1.0\linewidth]{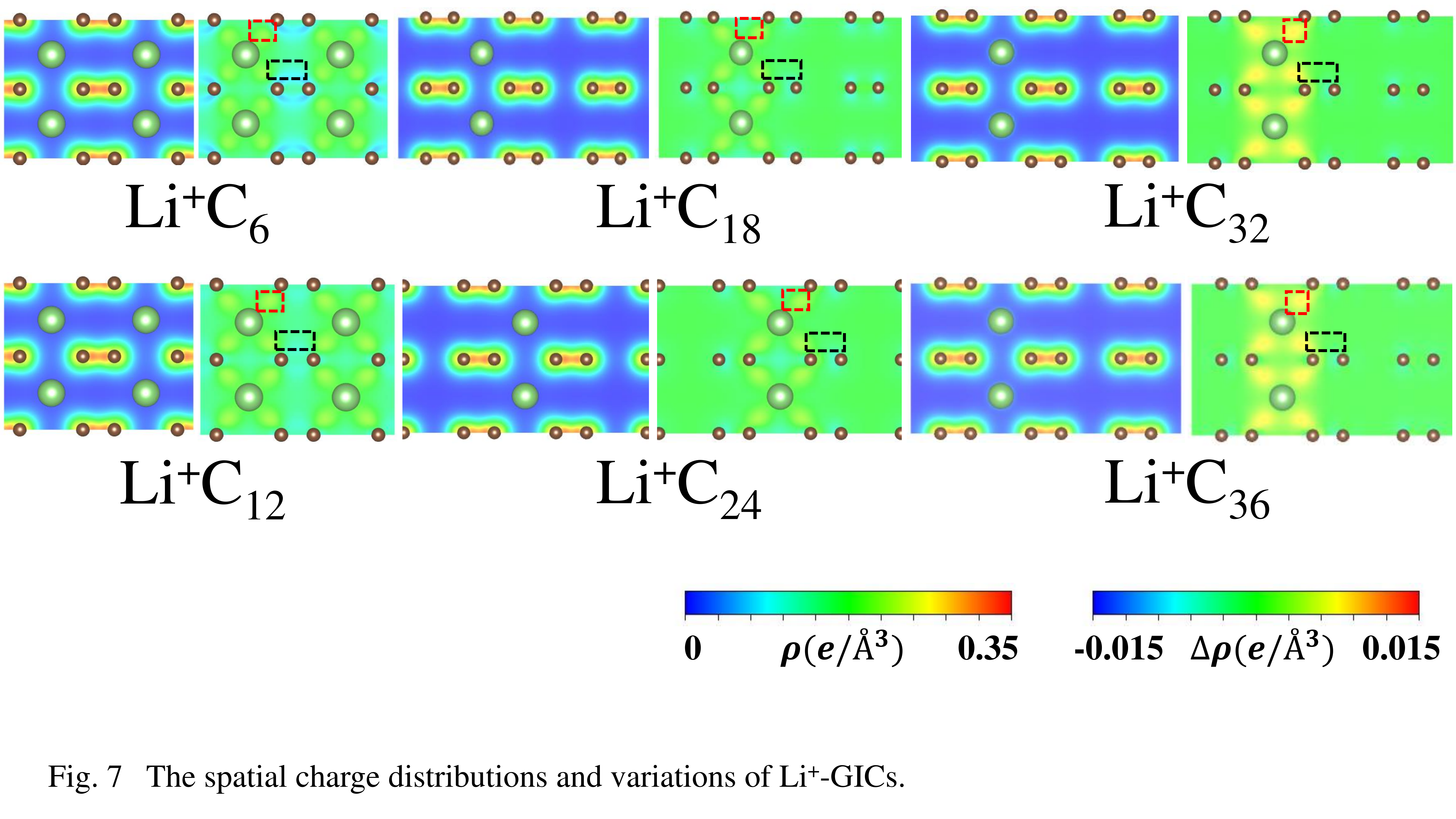}
\end{center}
\par
\end{figure}

\vfill
\eject
\end{document}